\documentclass[twocolumn]{jpsj2} 
\usepackage{color}
\newcommand{\LL}{$\Delta L/L_0${~}}

\title{Pressure Collapse of the Magnetic Ordering in MnSi via Thermal Expansion}

\author{Atsushi \textsc{Miyake}$^{1}$\thanks{atsushi.miyake@cea.fr}, Alain \textsc{Villaume}$^{1}$, Yoshinori \textsc{Haga}$^{1, 2}$, Georg \textsc{Knebel}$^{1}$, \\Bernard \textsc{Salce}$^1$, Gerard \textsc{Lapertot}$^{1}$, and Jacques \textsc{Flouquet}$^{1}$}

\inst{$^{1}$INAC/SPSMS, CEA-Grenoble, 17 rue des Martyrs, 38054, Grenoble, France \\
$^{2}$Advanced Science Research Center, Japan Atomic Energy Agency, Tokai, Ibaraki 319-1195, Japan\\
}

\abst{The itinerant quasi-ferromagnetic metal MnSi has been studied by detailed thermal expansion measurements under pressures and magnetic fields.
A sudden decrease of the volume at the critical pressure $P_c\sim1.6~\rm{GPa}$ has been observed and is in good agreement with the pressure variation of the volume fraction of the spiral magnetic ordering. 
This confirms that the magnetic order disappears by a first order phase transition.
The energy change estimated by the volume discontinuity on crossing $P_c$ is of similar order as the Zeeman energy of the transition from the spiral ground state to a polarized paramagnetic one under magnetic field. 
In contrast to the strong pressure dependence of the transition temperature, the characteristic fields are weakly pressure dependent, indicating that the strength of the ferromagnetic and the Dzyaloshinskii-Moriya interactions do not change drastically around $P_c$.
The evaluated results of the thermal expansion coefficient and the magnetostriction are analyzed thermodynamically.
The Sommerfeld coefficient of the linear temperature term of the specific heat is enhanced just below $P_c$.
The magnetic field-temperature phase diagrams in the ordered and paramagnetic phases are also compared.
Comparison is made with other heavy fermion compounds with first order phase transition at 0~K.
}

\kword{MnSi, spiral ordering, pressure, thermal expansion, ac-calorimetry,  first-order transition}

\begin{document}
\maketitle

\section{Introduction} 

High pressure is an ideal parameter to tune a strongly correlated electron systems (SCES) from one ordered state to another, e.g. from a magnetically ordered state to a paramagnetic Fermi liquid state.
Generally it is believed that such a zero temperature transition at a quantum critical point (QCP) is of second order.
Furthermore, close to the QCP the occurrence of new states of matter, such as pronounced non Fermi liquid (NFL) behavior and unconventional superconductivity, may be directly associated to the critical fluctuations due to the zero temperature quantum phase transition (QPT) \cite{HFroad, Stewart, Loheneysen}.
Here we are interested in the behavior of the thermal expansion through the QCP as function of pressure.
Generally a continuous volume $(V)$ change must occur at $T\rightarrow$0~K between a magnetically ordered state (antiferromagnetic (AF) or ferromagnetic (FM)) and a paramagnetic (PM) ground state.
The appearance of NFL properties is believed to be associated to the collapse of a characteristic energy scale of the critical fluctuations at the QCP \cite{Loheneysen}. 

The studies with temperature ($T$) and pressure ($P$)-tunings to QCP of SCES, such as itinerant magnetic metals and heavy fermion compounds, have revealed that the concept of a second order QPT can be invalid.
In the case of a FM-PM transition, for example, a first-order transition has been predicted theoretically and observed in the experiments \cite{pfleiderer_1997, Kirkpatrick_1999}
There are also examples of the first-order transition at $P_c$ for an AFM-PM transition, e.g. CeRh$_2$Si$_2$\cite{alain} and CeIn$_3$ \cite{Kawasaki}.   

In principle, one can distinguish a first order phase transition with a volume discontinuity ($\Delta V$) across $P_c$.
For a second order transition the volume thermal expansion coefficient ($\alpha_v = V^{-1}\partial V/\partial T$) will collapse at the QCP.
However, experimental evidences are not so easy to detect, because an intrinsic phase separation between the two states and/or the extrinsic pressure inhomogeneity may spread out any volume discontinuity close to the critical pressure.
Furthermore, also in the theory of three dimensional AF spin fluctuation even for a second order QCP, the N$\acute{\rm e}$el temperature may collapse strongly with a diverging Gr$\ddot{\rm u}$neisen parameter $\Omega_{T_{\rm N}} = - \partial \log T_{\rm N}/ \partial \log V$; that leads to a large thermal expansion coefficient $\alpha$, which is linked to $\Omega_{T_{\rm N}} \sim \alpha V_0/ C \kappa$ ($V_0$, $C$ and $\kappa$ are the molar volume, specific heat and compressibility) \cite{Zhu_2003}.
Thus, as in the case of a homogeneous first order transition at $T=0~\rm{K}$ (Clapeyron relation), the slope of $\partial T_{\rm N}/\partial P$ will be also infinite at the QCP.

In the present article, we focus on MnSi which has been intensively studied by macroscopic and microscopic experiments \cite{pfleiderer_1997, Yu_NMR, Uemura_muSR, Pfleiderer_2004, Thessieu_res, Pfleiderer_2001, Doiron_2003, Ishikawa_1976, Thessieu_1997, Fak_2005, Thessieu_NMR, Koyama_2000, Matsunaga_1982, Pedrazzini_2006, Ishikawa_1977, Pfleiderer_L, Petrova_2006, Stishov_2008, Franus_1984}.
It is well established that the magnetic transition gets first order above $P^{\ast}\sim$~1.2~GPa as function of temperature and vanishes at a critical pressure $P_c \sim$~1.5~GPa by a first order QPT \cite{pfleiderer_1997, Yu_NMR, Uemura_muSR}.
Further indications of the first order QPT are the appearance of intrinsic phase separation between the low pressure spiral spin structure \cite{Yu_NMR, Uemura_muSR} and another high pressure phase proposed to have partial magnetic order \cite{Pfleiderer_2004}.
Another singular point of MnSi is the observation of NFL properties as a $T^{3/2}$-dependence of the electrical resistivity on only in the vicinity of the QCP but at least up to $4.8~{\rm GPa}\sim3P_c$ \cite{Thessieu_res, Pfleiderer_2001, Doiron_2003, Pedrazzini_2006}, even though the partial ordering phase is suggested to be disappeared at $P \sim 2.1~{\rm GPa} \sim 1.5 P_c$ \cite{Pfleiderer_2004}.
Definitely, the new high pressure phase of MnSi does not look as a usual paramagnetic state.    

The low pressure helical (spiral) spin structure of MnSi with a long period of 180~\AA{} propagating along the [111] direction is well understood \cite{Ishikawa_1976, note}.
It has its origin in the so-called asymmetric Dzyaloshinskii-Moriya (DM) interaction due to the lack of inversion symmetry in its B20-type cubic structure.
The DM interaction precludes the establishment of a FM ground state \cite{Bak_1980, Kataoka_1981}.
However, a quite moderated field $H_c \sim$~0.6~T leads to restore a field polarized paramagnetic (PPM) phase below $T_c$.
On warming above $T_c$ this PPM appears under magnetic field through a crossover boundary ($H_m, T_m$) in the PM phase.
Previous magnetic susceptibility measurements under pressure up to 1.6~GPa showed a weak pressure dependence of $H_c$ or of the extrapolation ($H_m(T\rightarrow 0~\rm{K})$) for $P >P_c$ in contrast to the strong pressure dependence of $T_c$  \cite{Thessieu_1997}. 
This underlines clearly that the FM interaction itself is weakly pressure dependent.

Here we report on extensive thermal expansion measurements of MnSi under pressure and magnetic field. 
Furthermore we briefly present results of ac-calorimetry under pressure in zero magnetic field.
Special attention was given to increase the number of the measured pressure steps compared to the recent thermal expansion measurements under pressure via a neutron Larmor precession experiment \cite{Pfleiderer_L}. 
This allows a quantitative comparison with the other microscopic experiments such as  neutron scattering \cite {Fak_2005}, NMR \cite{Thessieu_NMR, Yu_NMR} and $\mu$SR \cite{Uemura_muSR}.
The pressure variation of the volume just below $P_c$ is well scaled with the volume fraction of the spiral phase proposed by NMR \cite{Thessieu_NMR, Yu_NMR} and $\mu$SR \cite{Uemura_muSR} experiments.
The fast volume collapse observed here near $P_c$ is the result of an inhomogeneous phase separation. 
If the first order transition at 0~K will be homogeneous, the volume discontinuity between the two phases will be estimated as $\Delta V/ V\sim 2\times10^{-4}$.
In terms of energy, the associated mechanical work $P\Delta V$ at $P_c$ corresponding to an energy of $\sim$0.4~K is quite comparable to the Zeeman energy $\mu_{\rm B} H$ to restore a PPM phase from the spiral phase at $H_c \sim 0.6~\rm T$, indicating the pressure insensitivity of FM and DM interactions.  

\section{Experimental details}


\subsection{Sample preparation}  
For the thermal expansion measurements, we used a sample of the same batch used in the previous neutron scattering measurements \cite{Fak_2005}. 
High-quality single crystals were prepared by Czochralski pulling from a stoichiometric melt of high-purity ($>99.995\% $) elements, using a radio-frequency heating and a cold copper crucible.
More details are reported in the literature \cite{Fak_2005}.

Small single crystals for the ac-calorimetry study have been grown by the zinc-flux method; the resistivity measurements in Ref.~\citen{Pedrazzini_2006} were realized on crystals of the same batch.

\subsection{Thermal expansion measurements} 
In order to measure the thermal expansion of MnSi under pressure, a strain gauge method is employed here.
The relative change of lattice constant can be detected by the change in the resistance of the gauge glued on sample.
It is possible to follow the magnetic anomaly on a relative extended pressure range.
In the interesting pressure range of 1~GPa~$<P<$~2~GPa near the suspected critical pressure $P_c\sim$~1.5~GPa, one can study the temperature evolution over a sufficiently large temperature window.
However, the temperature variation of the resistance of the gauge precludes measurements below 3~K.
A great advantage of this technique is that it is well suitable for measurements under pressure with the possibility to select an adequate pressure transmitting medium and to tune the pressure by fine steps close to $P_c$.
Of course, the difficulty is to ensure the coupling between the strain gauge and the deformation of the sample.

This technique under pressure is widely applied for heavy fermion compounds and gives quite reliable results \cite{alain, alain_URS, valentin_UGe2, Oomi_1997, Motoyama_URu2Si2}.
For example, the temperature variation of thermal expansion at ambient pressure is in good agreement with the result obtained by high accurate capacitor measurements.
We have succeeded to detect the anomaly corresponding to the phase transition of MnSi in agreement with previous works \cite{Pfleiderer_L} and also obtained that quite reproducible data are obtained during different pressure cycles.

We glued strain gauges on a single crystalline MnSi along $a$-axis and on a Si crystal, which is a reference having a negligibly small thermal expansion coefficient of less than $\sim$10$^{-7}$/K at least in the temperature range measured here\cite{silicon}. 
The relative change of the lattice constant, $\Delta L/L$, is measured by the so-called active dummy method and is proportional to the difference of resistance between the gauges on the sample and reference material, $\Delta R/R$.
The link between $\Delta R/R$ and $\Delta L/L$ was calibrated at 0.62~GPa by normalizing our data to the previous reports of the thermal expansion obtained from the neutron Larmor precession measurements,\cite{Pfleiderer_L} which gives absolute values. 
To detect the small change of the resistance, we used a Wheatstone bridge circuit.
More details about the experimental method is given elsewhere \cite{alain}.

Pressure was applied by using a two-layered hybrid piston cylinder type pressure cell filled with Daphne oil 7373 as pressure-transmitting medium. 
We determined the pressure by measuring the superconducting transition of Pb by ac-susceptibility.
For magnetostriction measurements, we applied magnetic fields along the $a$-axis with a conventional superconducting coil up to 2~T.
The calibrations under field, e.g. magnetoresistance of the gauge and magnetostriction of Si, were not taken into account, since they are negligibly smaller than the magnetostriction of MnSi.   
We performed two pressure cycles: the first measurements with increasing pressure up to 2~GPa (run 1), and the second measurements after releasing pressure to 1~GPa after run 1 and measuring with increasing pressure up to 1.7~GPa (run 2).

\subsection{ac-calorimetry measurements}
The temperature dependence of specific heat was measured by an ac-method under pressure and zero field.
Pressure was applied by a diamond-anvil cell filled with Ar as a pressure transmitting medium. 
The sample was heated using an Ar laser. 
By use of a mechanical chopper a quasi-sinusoidal excitation is transmitted to the sample via an optical fiber directly on the diamond. 
An Au/AuFe (0.07\%) thermocouple soldered to the sample served to measure the temperature oscillation $T_{\rm ac} = S_{\rm th} /V_{\rm th}$ of the sample, with  $S_{\rm th}$ and $V_{\rm th}$ being the thermopower and the measured voltage on the thermocouple. 
The measurements were performed at 720~Hz. 
The specific heat of the sample can be estimated by $C_{\rm ac} \propto \sin(\theta - \theta_0) / T_{\rm ac}$. 
An incertitude contains the origin of the phase $\theta_0$. 
Here, $\theta_0$ was chosen to reproduce the correct form of the temperature dependence of the ac specific heat at low temperature compared to an absolute measurement at zero pressure. Furthermore, the background contribution of the pressure cell is not known. This impedes a discussion of absolute values of the specific heat and furthermore to discuss the temperature dependences under pressure in more detail. Further details of the method and its limitation are given in Ref. \citen{Derr2006}. 

\section{Experimental results}

\subsection{Pressure-temperature phase diagram} 

\begin{figure}[tb]
\begin{center}
\includegraphics[width=0.9 \hsize,clip]{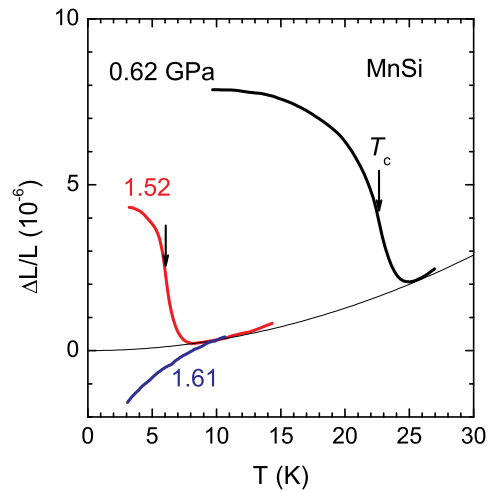}
\end{center}
\caption{(Color online) Temperature dependence of the relative change of lattice constant at 0.62, 1.52 and 1.62~GPa (run 1).
 The arrows are the evaluated $T_c$ by the extrema of temperature derivative (see Fig.~\ref{f2}).
The solid line indicates the conventional thermal expansion \LL $\sim 3.2\times 10^{-8} T^2$ reported in Ref.~\citen{Pfleiderer_L}.}
\label{f1}
\end{figure}

In a first run,  we measured the temperature dependence of the linear thermal expansion \LL along $a$-axis at 11 pressures up to 2~GPa in zero magnetic field. 
Fig.~\ref{f1} presents $\Delta L (T)/ L_0$ at selected pressures of 0.62, 1.52 and 1.61~GPa.
\LL is generally the sum of the lattice, electronic and magnetic contributions.
In an itinerant magnetic systems, the latter two-terms can not be decoupled.
At 0.62 and 1.52~GPa, \LL follows roughly a $T^2$-dependence above the temperature of its minimum, which corresponds to the conventional lattice thermal expansion.
A solid line in the figure is the reported conventional thermal expansion \cite{Pfleiderer_L}.
The slight deviations of our data from the line are due to the pressure variation of the $T^2$ term and discussed later.
On further cooling a negative thermal expansion is observed.
It implies the temperature evolution of the spontaneous magnetostriction which linked to the formation of the sublattice magnetization.
With increasing pressure, the spontaneous magnetostriction decrease. 
At 1.61~GPa, there is no negative thermal expansion at least down to 3~K, indicating the suppression of the spiral magnetic ordered phase.

\begin{figure}[tb]
\begin{center}
\includegraphics[width=0.9 \hsize,clip]{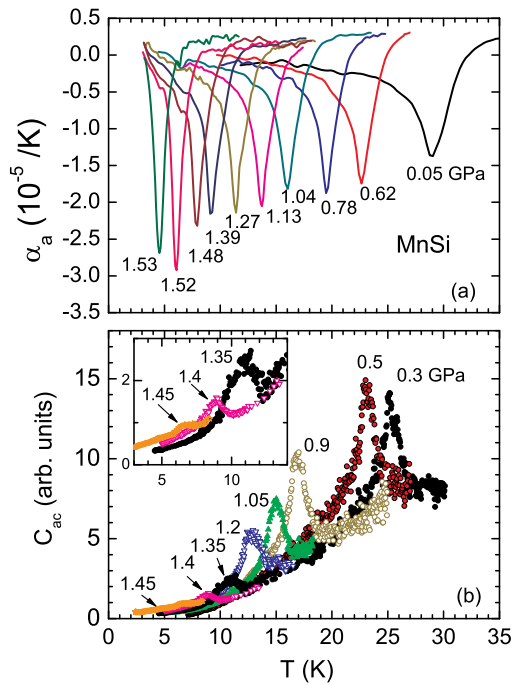}
\end{center}
\caption{(Color online) Temperature dependence of (a) the linear thermal expansion coefficient along the $a$-axis $\alpha_a$ (run 1) and (b) the specific heat obtained by the ac-method $C_{\rm ac}$ for several pressures.
The inset in (b) shows pressure evolution of specific heat anomalies near $P_c$.}
\label{f2}
\end{figure}

\begin{figure}[htb]
\begin{center}
\includegraphics[width=0.9 \hsize,clip]{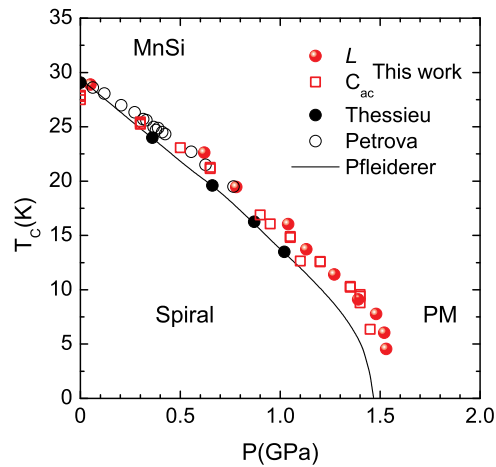}
\end{center}
\caption{(Color online) Pressure temperature phase diagram of MnSi. 
Our results indicated by the full circles and open squares are obtained by the thermal expansion and specific measurements.
The data determined by the other probes, ac-susceptibility (solid line cited from Ref.~\citen{pfleiderer_1997} and black open circle cited from Ref.~\citen{Petrova_2006}) and resistivity (black circle cited from Ref.~\citen{Thessieu_res}) are also shown for comparison. }
\label{PD}
\end{figure}

The thermal expansion coefficient along $a$-axis $\alpha_a \equiv \partial(\Delta L/L_0)/\partial T$  is plotted in Fig.~\ref{f2}(a).
$T_c$ is defined at the extrema of $\alpha_a$ and indicated by arrows in Fig.~\ref{f1}. 
Fig.~\ref{f2}(b) represents the temperature dependence of the specific heat derived from ac-calorimetry on a single crystal from different batches.
Qualitatively, the agreement between thermal expansion coefficient and specific heat is excellent (see also Fig.~\ref{PD}).
Above 1.3~GPa, the specific heat anomaly at the magnetic transition is strongly suppressed, in contrast to the slight enhancement of the anomaly of $\alpha$ with pressure.
 For $P > 1.45$~GPa, no anomaly is observed in any specific heat experiment above 2~K. 

In Fig.~\ref{PD}, we summarize the pressure-temperature phase diagram determined by the peak of $\alpha_a(T)$ and specific heat at zero field together with previously reported data \cite{pfleiderer_1997, Thessieu_res, Petrova_2006}.
Compared to our results, the reported value of $P_c$=1.46~GPa in Ref.~\citen{pfleiderer_1997} and \citen{Thessieu_res} is slightly lower. 
This discrepancy may be due to the difference in the pressure inhomogeneity due to different pressure media.
Another origin might come from the different quality of the MnSi crystals. 
The resistivity measurements on sample from the same batch used in thermal expansion measurements gives a residual resistivity ratio near 40.

\subsection{First-order transition at $P_c$; volume discontinuity}

In a second run, we performed more detailed pressure steps. 
Fifteen different pressures are measured in the pressure range between 1~GPa and 1.7~GPa in order to obtain more detailed insights in the behavior of the system in the vicinity of $P_c$. 
Usually, a solid contracts on cooling above the the onset of the magnetic transition.
This high temperature part can be fitted by a $a'T^2$ term describing the lattice contribution.
In order to obtain the spontaneous magnetostriction, the $T^2$-term is subtracted from the total thermal dilatation $\Delta L/ L_0$.
With this procedure in our experiment, the evaluated $a'$ varies with $P$: e.g. $a'$  for $P=$ 0, 1, and 1.5~GPa are approximately 3, 5 and 7$\times 10^{-8}~{\rm K}^{-2}$.

\begin{figure}[!h]
\begin{center}
\includegraphics[width=0.9 \hsize,clip]{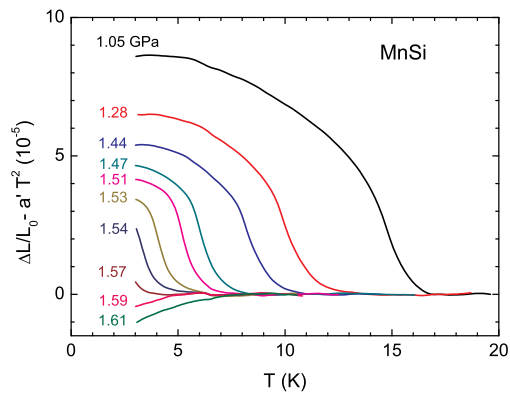}
\end{center}
\caption{(Color online) Temperature dependence of the spontaneous magnetostriction for several pressures (run 2). The data were obtained by subtracting the $T^2$-term which has been fitted for every pressure above $T_{\rm{min}}$ where \LL has a minimum (see the text). }
\label{dLvsT}
\end{figure}

\begin{figure}[tb]
\begin{center}
\includegraphics[width=0.9 \hsize,clip]{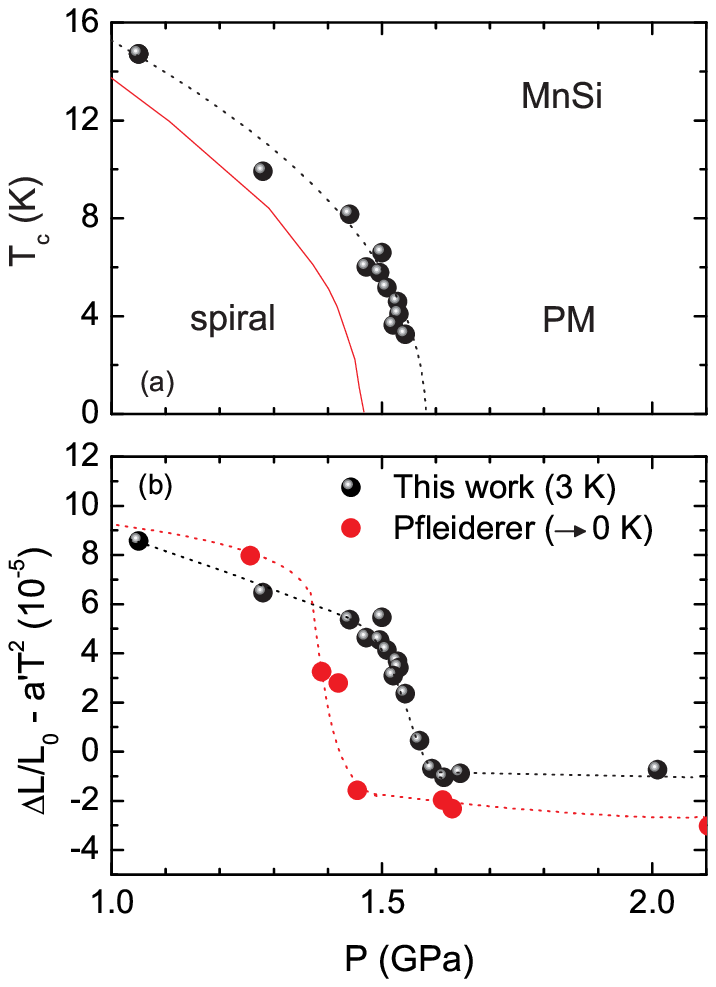}
\end{center}
\caption{(Color online) Pressure variation of (a) the ordering temperature $T_c$ and (b) the spontaneous linear thermal expansion taken at 3~K (see data of Fig.\ref{dLvsT}). The results from the neutron Larmor precession \cite{Pfleiderer_L} are also plotted in both figures: the solid (red) line in the (a) and red circles in the (b). The dotted lines are guides for eyes.}
\label{PD&a&dV}
\end{figure}

Fig.~\ref{dLvsT} presents the temperature dependence of the spontaneous magnetostriction $\Delta L/L_0 -a'T^2$ at various pressures above 1~GPa.
With increasing pressure, the spontaneous magnetostriction decreases gradually up to $\sim$ 1.4~GPa and is suddenly suppressed on approaching $P_c$ (see also Fig.~\ref{PD&a&dV} (b)).
Across the critical pressure $P_c$, it changes sign and gets negative at low temperature and shows weak pressure variation above $P_c$, in agreement with the neutron Larmor precession measurements \cite{Pfleiderer_L}.


In Fig.~\ref{PD&a&dV}, we plot the pressure dependence of $T_c$ and of the spontaneous magnetostriction obtained at 3~K in Fig.~\ref{dLvsT}.
From our results, $P_c$ is determined as $P \sim$~1.58~GPa within experimental uncertainty.
The volume discontinuity expected by the nature of the first-order phase transition is clearly observed from $P\sim 1.5$~GPa  to $P_c=$~1.58~GPa.
The pressure width $\Delta P\sim 0.1$~GPa of the volume discontinuity is almost similar to the previous results by extrapolating to 0~K  as shown in Fig.~\ref{PD&a&dV}(b) \cite{Pfleiderer_L}. 
As discussed later, the discontinuity corresponds to the phase separation between spiral and PM phase.

\begin{figure}[tb]
\begin{center}
\includegraphics[width=0.9 \hsize,clip]{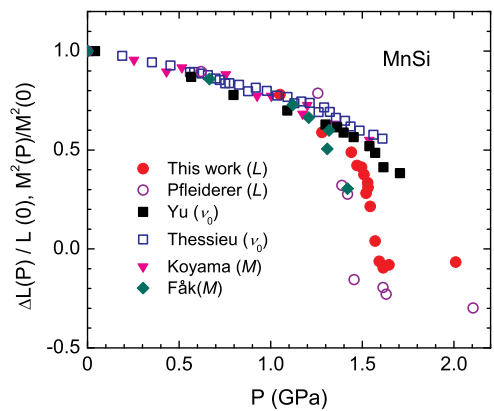}
\end{center}
\caption{(Color online) Pressure dependence of the the spontaneous magnetostriction $\Delta L (P)/L(0)$ (from Fig.~\ref{PD&a&dV}(b)) and the square of the magnetization $M^2(P)/M^2(0)$.
All data are normalized at ambient pressure.
Closed and open circles are the results of $\Delta L(P)/ L(0)$ in this work and in Ref.~\citen{Pfleiderer_L}, respectively.
The $M^2(P)/M^2(0)$ are obtained from the NMR resonant frequency ($\nu_0$) (squares) \cite{Yu_NMR, Thessieu_NMR}, bulk magnetization measurement (triangles) \cite{Koyama_2000} and neutron scattering measurement (diamonds) \cite{Fak_2005}, respectively.  }
\label{LM-P}
\end{figure}

The magneto-volume effect in a weak itinerant ferromagnet is generally expressed by the square of the magnetization $M$, $\Delta L /L_0 \propto M^2$ \cite{mor1980}.
In order to compare to the magnetization, we replot the spontaneous magnetostriction normalized at zero pressure as shown in Fig.~\ref{LM-P}.
The $M^2$ plotted in the figure are taken from the bulk magnetization\cite{Koyama_2000}, NMR resonant frequency \cite{Yu_NMR, Thessieu_NMR} and integrated neutron diffraction intensity \cite{Fak_2005}.
All results are in good agreement up to at least $P\sim 1.2$~GPa.
Above 1.2~GPa,  \LL and the sublattice magnetization from the neutron scattering\cite{Fak_2005} are strongly suppressed, although $M^2$ obtained by the bulk magnetization and NMR resonant frequency gradually decreases and still have finite values above $P_c$.
This is in agreement with the statements of NMR \cite{Yu_NMR} and $\mu$SR \cite{Uemura_muSR} experiments.
A phase separation occurs in the pressure range $P^{\ast} < P < P_c$.
The volume fraction of the spiral phase starts to be suppressed above $P^{\ast}$.

\begin{figure}[tb]
\begin{center}
\includegraphics[width=0.9 \hsize,clip]{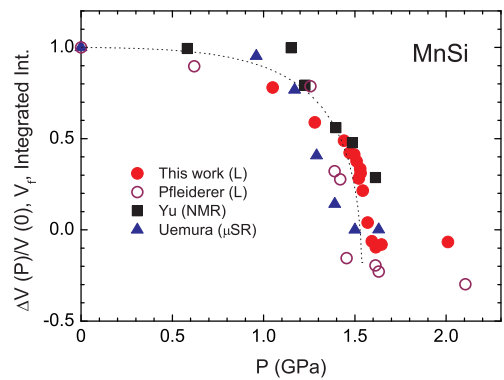}
\end{center}
\caption{(Color online) Comparison of the volume change $\Delta V/ V$ (circles, after Fig.~\ref{PD&a&dV}(b)), the integrated NMR spectrum intensity (squares) \cite{Yu_NMR} and the volume fraction $V_f$ obtained by $\mu$SR,  (triangles) \cite{Uemura_muSR} as a function of pressure. 
All  data are normalized at ambient pressure. 
The line is a guide for eyes.
  }
\label{LVf-P}
\end{figure}

We also compare the spontaneous magnetostriction with the volume fraction of spiral phase obtained by the $\mu$SR \cite{Uemura_muSR} and from the integrated intensity of NMR spectrum \cite{Yu_NMR} in Fig.~\ref{LVf-P}.
The results excellently agree with each other except for the small difference of the critical pressures.
The pressure width of the first-order transition is considered to arise from an extrinsic pressure gradient and/or the intrinsic phase separation.

\begin{figure}[tb]
\begin{center}
\includegraphics[width=0.9 \hsize,clip]{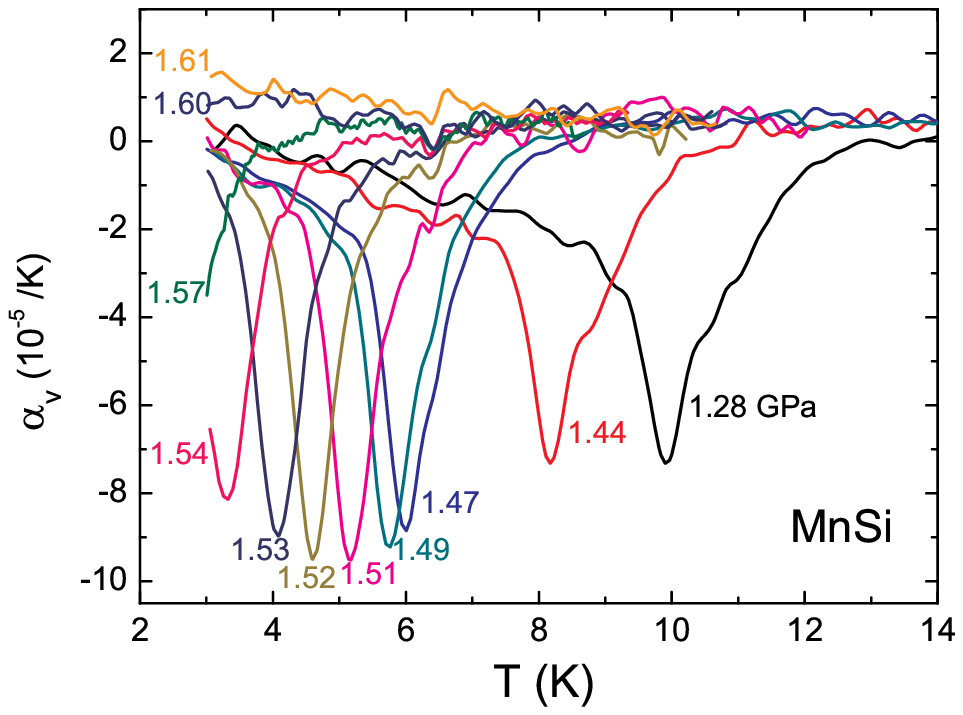}
\end{center}
\caption{(Color online) Temperature dependence of the volume thermal expansion coefficient $\alpha_v=3\alpha_a$at several pressures and zero field (run 2).}
\label{Alpha}
\end{figure} 

Without any phase separation, a step-like volume discontinuity at $P_c$ is expected.
Assuming that the volume is expressed as $V=V_{\rm S}f_{\rm S}+V_{\rm p.o}(1-f_{\rm S})$, where $V_{\rm S (p.o)}$ is the volume of the contribution from the spiral order (S) and the partial order (p.o), and $f_{\rm S}$ is the volume fraction of spiral ordered phase, the volume change is obtained as $\Delta V= V_{\rm S}-V= (V_{\rm S}-V_{\rm p.o})(1-f_{\rm S})$.
We obtained $(V_{\rm S}-V_{\rm p.o})/V_{\rm S} \sim 2\times10^{-4}$.  
In Fig.~\ref{Alpha},  the pressure variation of the temperature dependence of volume thermal expansion coefficient $\alpha_v (= 3\alpha_a)$ found in run~2 are shown.
With increasing pressure up to $P\sim$1.5~GPa, an increase of the deep minimum anomaly of $\alpha$ at $T_c$ associated with a sharpening is in agreement with the Ehrenfest's principle.
Above 1.51~GPa, a strong decrease in the amplitude of the anomaly with $P$ is clearly seen and strongly related to the collapse of $f_{\rm{S}}$ to zero above $P_c$.
These behaviors agree excellently with the pressure dependence of the spontaneous magnetostriction close to $P_c$ as shown in Fig.~\ref{PD&a&dV}.
From our experiments, the phase separation occurs in a pressure range of 0.1~GPa, which is smaller than the previous reports of $\Delta P = P_c - P^{\ast} \sim 0.3$~GPa \cite{Yu_NMR, Uemura_muSR}.
$P^{\ast}$ is considered to be less sensitive than $P_c$, since it described the beginning of the phase separation.
Basically it occurs at a pressure where $\partial T_c/\partial P$ is not so high by comparison to that at $P_c$. 
However, $P^{\ast}$ seems to depend on the sample condition. 
For example, the two independent ac-susceptibility measurements with different pressure mediums and samples give the different values of $P^{\ast}$ \cite{pfleiderer_1997, Petrova_2006}. 



\subsection{$H-T$ phase diagram \label{BT-phase diagram}}

\begin{figure}[tb]
\begin{center}
\includegraphics[width=0.9 \hsize,clip]{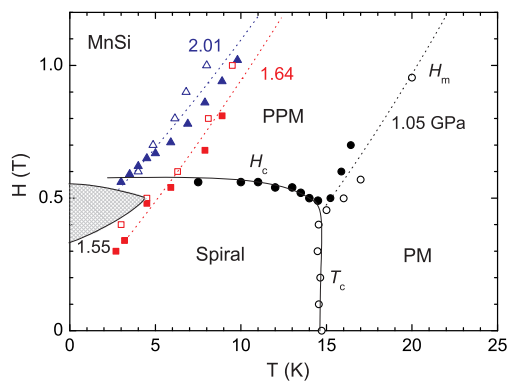}
\caption{(Color online) Temperature-field phase diagram of MnSi at pressures of 1.05 (circle), 1.64 (square) and 2.01~GPa (triangle). 
The guides of solid and dotted lines indicate the phase boundaries ($H_c$ and $T_c$) and a crossover ($H_m$), respectively. 
The open and closed symbols indicate the data determined from the magnetostriction and the thermal expansion measurements.
The shaded region indicates the field re-entrant phase at 1.55~GPa reported by the ac-susceptibility measurements \cite{Thessieu_1997}.}
\label{BT-PD}
\end{center}
\end{figure}

The $H-T$ phase diagrams at pressures of 1.05~GPa ($<P_c$), 1.64~GPa ($\sim P_c$) and 2.01~GPa ($>P_c$) obtained by the thermal expansion and the magnetostriction measurements are presented in Fig.~\ref{BT-PD}. 
$H_c$ and $H_m (T_m)$ are the transition field restoring from spiral and PPM phases and crossover field (temperature) between PPM and PM, respectively.

\begin{figure}[tb]
\begin{center}
\includegraphics[width=0.9 \hsize,clip]{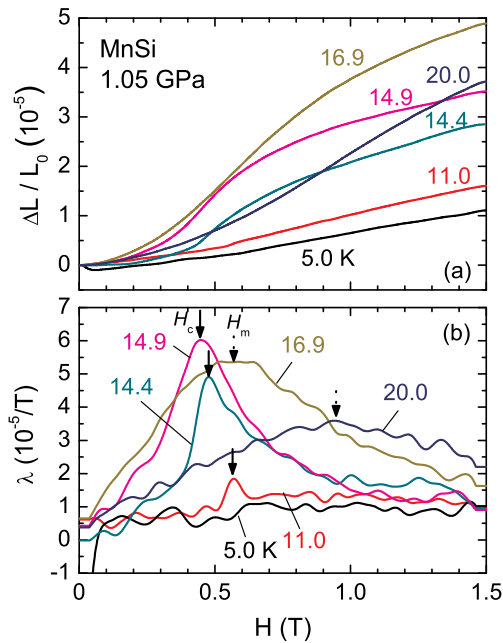}
\caption{(Color online) Field variation of (a) the relative change of lattice constant \LL and (b) their field derivative $\lambda$ at a constant pressure of 1.05~GPa and several temperatures.
The solid and dotted arrows indicate the transition field between the spiral and PPM, $H_c$, and the crossover from PM to PPM, $H_{\rm m}$.}
\label{L-B(1GPa)}
\end{center}
\end{figure}

Fig.~\ref{L-B(1GPa)} shows field dependence of the relative change of the lattice constant $\Delta L/ L_0$ and $\lambda \equiv  \partial (L/L_0)/\partial H$ at a pressure of $P=$~1.05~GPa~$<P_c$. 
At higher temperature an anomaly around $H\sim 0.5~T$ is observed, which is in agreement with the transition field from the spiral order to a polarized paramagnetic (PPM) regime \cite{Thessieu_1997}. 
This phase boundary $H_c$ is very ambiguous at low temperature and becomes well defined above $T \sim$~10~K.  
This can  be seen in Fig.~\ref{L-B(1GPa)}(b), which is in agreement with the weak pressure dependence of the magnetization \cite{Koyama_2000} according to the Maxwell relation. 
The magnetostriction is more pronounced on approaching to $T_c~\sim$~15~K. 
For higher temperatures $T > T_c$ the peak is broadened and no clear phase boundary can be drawn.
A broad peak of $\lambda(H)$ denoted as $H_m$ in Fig.~\ref{L-B(1GPa)}(b) is observed under magnetic field, as has been already observed by susceptibility \cite{Thessieu_1997}: that corresponds to the crossover from PPM to PM.

\begin{figure}[tb]
\begin{center}
\includegraphics[width=0.9 \hsize,clip]{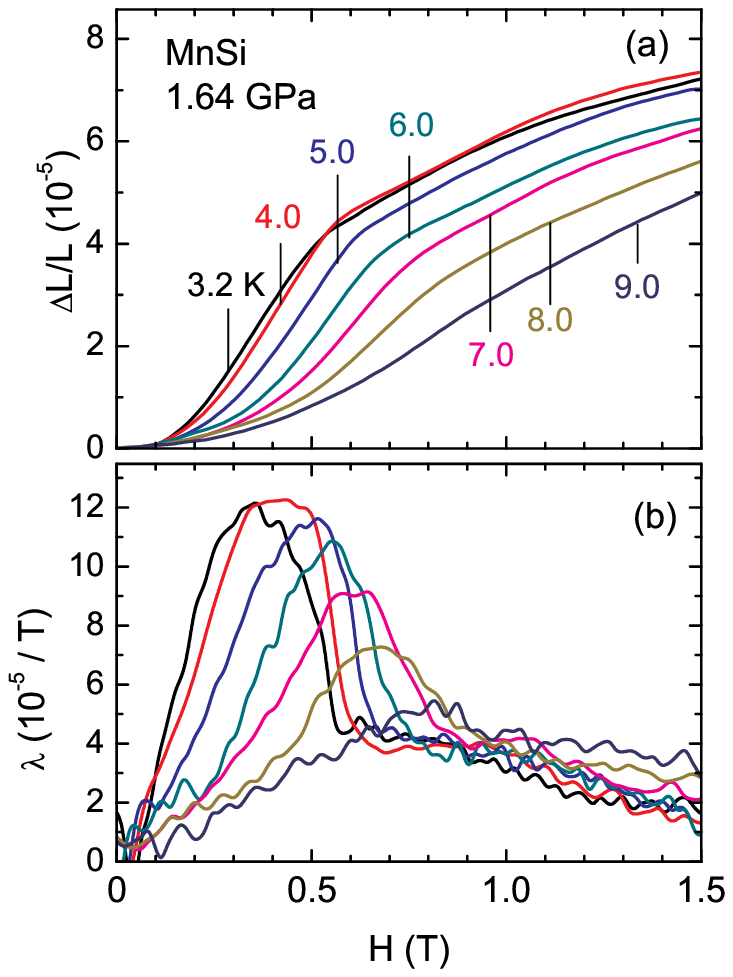}
\caption{(Color online) Field variation of (a) the relative change of lattice constant \LL and (b) $\lambda$ at a constant pressure of 1.64~GPa and several temperatures. }
\label{L-B(16kbar)}
\end{center}
\end{figure}

Just above $P_c$ at a pressure of 1.64~GPa, a sharp kink in the magnetostriction is observed around 0.5~T for $T=3.2$~K as shown in Fig.~\ref{L-B(16kbar)}.
The peak in $\lambda$ shifts to higher fields and broadens with increasing temperature, as already observed at $H_m$ at 1.05~GPa in the Fig.~\ref{L-B(1GPa)}.
A field re-entrant magnetic phase shown in the Fig.~\ref{BT-PD} at low temperature and at 1.55~GPa close to $P_c$ was clearly observed in the field variation of susceptibility \cite{Thessieu_1997}.
Our magnetostriction measurements failed to detect this re-entrant phase presumably due to the temperature limitation in our experiment.
As shown in Fig.~\ref{L-B(16kbar)}, however, the hight of peak in $\lambda(H)$ is almost constant or slightly increases from 3.2 to 4.0~K, in contrast to the strong suppression at higher temperature.
It seems to be consistent with the susceptibility measurements at 1.55~GPa, where an enhancement of the susceptibility peak with increasing temperature on approaching the tri-critical point among the re-entrant helical or spiral, PPM and PM phases has been observed \cite{Thessieu_1997}. 

\begin{figure}[tb]
\begin{center}
\includegraphics[width=0.9 \hsize,clip]{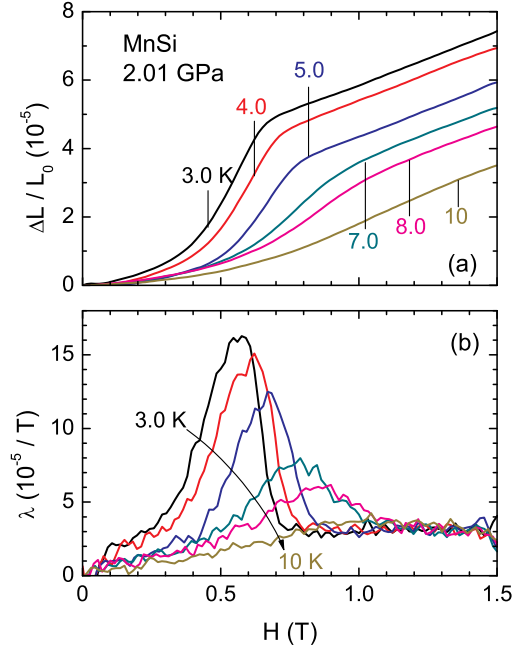}
\caption{(Color online) Field variation of (a) the relative change of lattice constant \LL and (b) $\lambda$ at a constant pressure of 2.01~GPa and several temperatures.}
\label{L-B(2GPa)}
\end{center}
\end{figure}

Well above $P_c$ (Fig.~\ref{L-B(2GPa)}), the peak in $\lambda$ is more pronounced, although the temperature evolution of $\lambda$ is quite similar to that at 1.64~GPa.
It might be interesting to note that $\lambda$ at 2.01~GPa and for fields higher than $H\sim 1$~T is almost constant and insensitive to temperature.

\section{Discussions}

We compare in Table~\ref{mechanical work} the volume discontinuity of different systems believed to have a first order transition.
Quite similar values of $\Delta V/ V$ are found for the four intermetallic systems which have also quite similar values of $P_c$ despite the fact that they involve different ground states at $P_c$.
For CeRh$_2$Si$_2$, the ground state switch occurs between AF and mixed valence paramagnetic phase, for UGe$_2$ between two FM states (FM2 and FM1) and for URu$_2$Si$_2$ between hidden order (HO) and AF phases. 
It is remarkable that, for the four systems, the idealized jump of the sublattice magnetization at the critical pressures have the same magnitude: 0.3~$\mu_{\rm B}$ for MnSi, 0.4~$\mu_{\rm B}$ for CeRh$_2$Si$_2$, 0.4~$\mu_{\rm B}$ for UGe$_2$ and 0.3~$\mu_{\rm B}$ for URu$_2$Si$_2$.	 
By the analogy of the solid-liquid phase transition of $^3$He, the compounds in Table~\ref{mechanical work} cannot be classified by a weak first-order transition.
For $^3$He, $\Delta V/V$ is $\sim 5\times 10^{-2}$, but $P$ is $\sim 3.4$~MPa which is basically three order of magnitude lower than $P_c$ of the other tabulated materials. 
For the reported heavy fermion compounds the mechanical work $P\Delta V/V$ at the critical pressure corresponding to the energy change across the phase transition in all materials is due to a volume change $\Delta V / V$ of $\sim 10^{-4}$.
Assuming that $P\Delta V$ is equivalent to the energy, we get $T\sim$~0.4~K for MnSi, from the spiral to the partial ordered phase transition, which is comparable to the field collapsing the spiral order $H_c\sim$~0.6~T.   
As discussed previously \cite{Thessieu_1997} and in subsection \ref{BT-phase diagram}, the characteristic fields, $H_{c}$ or $H_m$ extrapolated to $T\rightarrow0~{\rm K}$, are weakly pressure dependent, in contrast to the strong pressure variation of $T_c$.
For example, $H_c(0) \sim0.55$~T at $P=$~1.05~GPa is near the extrapolation of $H_m(0)\sim$ 0.4~T for $P=2.01~{\rm GPa}$ as seen in Fig.~\ref{BT-PD}.
By analogy with the re-entrant phase at $P = 1.55$~GPa \cite{Thessieu_1997}, which we failed to detect due to the experimental temperature limit ($T>3~\rm K$) at $P = 1.64$~GPa, $H_c(0)$ around $P_c$ have also similar value of 0.6~T. 
Thus the FM and DM interactions appear robustly above $P_c$.


\begin{table}[tbp]
\label{t1}
\caption{Comparison of the relative volume change,  $\Delta V/V$, at the first-order phase transition in several materials.
The type of phase transition and critical pressure are shown for comparison. 
The IV and HO indicate the intermediated valence and hidden order phase.
For UGe$_2$, the two different Ising-type of FM with different moments are indicated as FM1 and 2. 
A typical example of a first-order transition from solid to liquid for $^3$He is also shown.
\\}

\begin{tabular}{l c c c }

\hline
&  Phases &$P$ & $\Delta V/V $  \\ 
\hline
MnSi (This work) & FM-NFL      & 1.5~GPa  &   $\sim2\times 10^{-4}$      \\
CeRh$_2$Si$_2$\cite{alain}   &  AFM-IV   &  1.2~GPa &   $\sim1\times 10^{-4}$       \\
UGe$_2$\cite{valentin_UGe2}          & FM2-FM1   & 1.2~GPa  & $\sim1.5\times 10^{-4}$     \\
URu$_2$Si$_2$\cite{Motoyama_URu2Si2}    & HO-AF      &  $\sim $0.5~GPa &  6$\times 10^{-5}$     \\
$^3$He & solid-liquid &  3.4~MPa &    $\sim5\times 10^{-2}$\\
\hline

\label{mechanical work}
\end{tabular}

\end{table}

\begin{figure}[tb]
\begin{center}
\includegraphics[width=0.9 \hsize,clip]{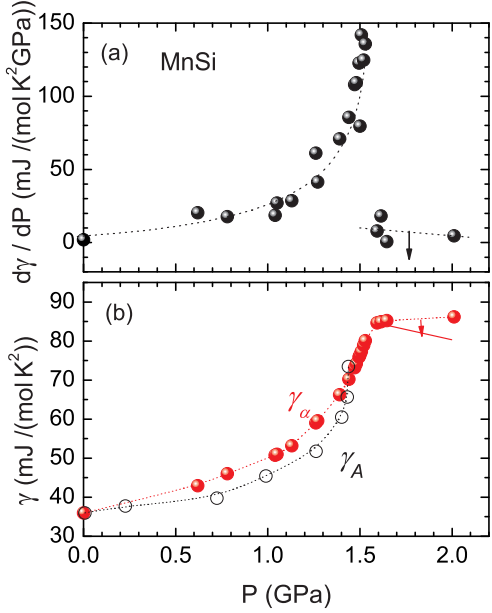}
\end{center}
\caption{(Color online) Pressure dependence of (a)$\partial \gamma / \partial P$ and (b)$\gamma$ evaluating from the volume thermal expansion coefficient (solid circle), $\gamma_{\alpha}$,  and from the $T^2$-coefficient of resistivity (open circle), $\gamma_{A}$ \cite{Thessieu_res}.
As shown in arrows, above $P_c$, $\partial \gamma / \partial P$ should be negative, resulting in the negative pressure dependence of $\gamma$ (solid line in (b)).
The details are discussed in the text. 
The dotted lines are guides for eyes.}
\label{dg/dp}
\end{figure}

Next, we try to evaluate how the low temperature electronic excitations change under pressure, i.e. what is the pressure evolution of the effective mass.
This can be achieved when the thermal expansion reaches its low temperature regime: $\Delta L/L_0$ varies as $T^2$, and hence the $\alpha_v \propto T$.
According to the Maxwell relation,
\begin{equation}
\frac{\partial V}{\partial T} = -\frac{\partial S}{\partial P} = -T\frac{\partial \gamma}{\partial P}.
\label{Max_dV/dT}
\end{equation} 
Then, we obtain the pressure derivative of the Sommerfeld coefficient $\gamma$ of the linear temperature term of the specific heat from the coefficient of $T$-term of $\alpha_v$ from Fig.~\ref{Alpha},
\begin{equation}
\frac{\partial \gamma}{\partial P} = -V_0\frac{\alpha_v}{{T}}.
\end{equation} 
A derivation of the low temperature regime can not be achieved near $P_c$.
However, it can be derived for $P <$~1.51~GPa as $T_c \sim$~5~K, and attempts can be made above $P_c\sim$~1.6~GPa.
Fig.~\ref{dg/dp}(a) represents the derived slope of $\partial \gamma/ \partial P$ ; there are a steep increase of $\partial \gamma/ \partial P$ below $P_c$ and a weak variation of that above $P_c$. 
That leads to the pressure variation of $\gamma$ shown in Fig.~\ref{dg/dp}(b).
Of course, one can argue on the accuracy in $\partial \gamma/\partial P$ above $P_c$.
Without any subtraction of any $T^2$ contribution to the lattice deformation, in our experiment the crossing of $T_c$ leads to the expansion of the lattice on cooling.
Above $P_c$, a continuous contraction is observed corresponding to the classical behavior that band width as well as the related Fermi temperature increase with pressure (see Fig.~\ref{f1}).
Above $P_c$ and at low temperature, $\alpha$ must have a positive temperature dependence, leading to the negative $\partial \gamma/\partial P$ above $P_c$.
In heavy fermion compounds an estimation of the pressure variation of $\gamma$ is often given through the $P$ dependence of the $A$-coefficient of the $AT^2$ resistivity term assuming the validity of the Kadowaki-Woods relation $A\sim \gamma^2$.
Our derived values of $\gamma(P)$ can be scaled rather well with $\sqrt{A}$\cite{Thessieu_res} at least up to 1.4~GPa.
Of course above $P_c$, the collapse of the Fermi liquid law in the resistivity prevents any comparison.


It is also interesting to remark that (i) the predicted $\gamma(P)$ of MnSi is quite comparable to that found in CeRh$_2$Si$_2$ \cite{Graft_1997} and even in UGe$_2$ \cite{Tateiwa_2001} and (ii) the NFL behaviors observed in MnSi above $P_c$ has been also found for URu$_2$Si$_2$ at the first order HO-AF transition \cite{Hassinger_2008} and also for CeRh$_2$Si$_2$ at the first order transition not between AF and PM but between the two AF ground states \cite{BoursierThesis}.


The different ground states achieved in MnSi under pressure are dictated by the hierarchies of the ferromagnetic exchange, DM interaction and local anisotropy fixing the direction of propagation vector.
The image of the previous experiment is that the FM interaction is weakly $P$ invariant as well as the DM interaction (weak pressure dependence of $H_c(0)$ below $P_c$ and of $H_m(0)$ above $P_c$) \cite{Thessieu_1997}.
Thus the drastic pressure effects may be some changes in the anisotropy strongly related to the local nature of the magnetism originated from the Mn-ion.
Here, it is worthwhile to notice that at ambient pressure, MnSi appears as a mixed valent compound \cite{Carbone_2006}.
Thus as it happens for 4$f$- or 5$f$-based heavy fermion systems \cite{JF_JPSJ}, tiny change of the valence may have drastic effects in the anisotropy \cite{HFroad}.

The unusual phase above $P_c$ is still not identified.
It was reported from neutron scattering experiments that a partial magnetic order seems to occur with the comparison of short range orders in liquid crystals \cite{Pfleiderer_2004}.
Recently, the switch from single wave vector (spiral) ordering to an amorphous skymiron ground state \cite{Rossler_2006}, to new crystalline phases (equivalent to blue phases in liquid crystal) \cite{Fischer_2008} or to superposition of distinct spin spirals \cite{Binz_2006} have been suggested.
There is clearly a necessity to invoke topological defects or high sensitivity to disorders to try to elucidate the NFL behavior detected in transport measurements above $P_c$ \cite{Pfleiderer_2001, Doiron_2003}.
The unusual high sensitivity of MnSi with its chiral DM interaction leads already to the proposal of an easy formation of extended localized defects above $T_c$ with the interesting description of extended localized defects \cite{Buzdin}.

As pointed out, the insufficient precision of our experiments and previous works \cite{Pfleiderer_L} do not allow to derive accurate temperature dependence of the thermal expansion at low temperature and above $P_c$.
In both cases, no signature of the phase transition at $T_0$ emerges in $\alpha$ has been detected.
In our experiments, it has been found that just above $T_c$ the fit of the $\Delta L/ L_0$ data by a $a'T^2$ term requires a pressure variation of $a'$.
In the previous Larmor precession experiments, a constant value of $a'$ was taken \cite{Pfleiderer_L} .
When $T_c$ decreases significantly, this term cannot be a pure phonon contribution as the lattice thermal expansion related to the phonon specific heat ($C \propto T^3$) via the Gr$\ddot{\rm u}$neisen parameter of the Debye temperature must correspond to a $T^4$ dependence of the lattice contraction.
Our guess is that in this report the derived extra characteristic energy $T_{\rm TE}$ corresponds to the regime where the thermal dilatation becomes dominated by the electronic degree of freedom.

An important challenge is to improve the accuracy in the thermal expansion experiments under pressure.
In addition, the difficulties of every high pressure experiments are to evaluate the role of pressure inhomogeneity in the pressure chamber through the pressure transmitting medium.
A pressure distribution near 0.05~GPa is often estimated with the Daphne oil used here \cite{Jaccard_2007}.
Fluorinert (depending on the constituent of the mixture) often used for neutron pressure measurements and the other organic liquid can induce even higher pressure inhomogeneity \cite{Sidorov_2005}.
A new generation of experiments must be realized with an optimization of the hydrostaticy, for example, transport measurements to investigate the unusual phase for $P > P_c$ in diamond anvil cell with He transmitted medium.

\section{Conclusion}

Extensive thermal expansion measurements of MnSi under pressure give the value of the volume discontinuity $\Delta V/ V\sim2\times 10^{-4}$ across $P_c$, which is well scaled by the volume fraction of spiral ordered state derived by microscopic measurements, such as NMR and $\mu$SR.
The energy associated with the volume discontinuity, $P\Delta V$, is comparable to the Zeeman energy required to destroy the chiral order.
The similar volume discontinuity associated with comparable drop of the sublattice magnetization was found for the heavy fermion systems showing the first order transition between quite variety of  ground states, such as CeRh$_2$Si$_2$, UGe$_2$ and URu$_2$Si$_2$.

The ground state above $P_c$ is still under debate.
The FM exchange energy is quite stable under pressure as well as the DM interaction, i.e. the weak $P$ variation $H_c(0)$ and $H_m(0)$. 
Our proposal is that the main pressure driving change originates from the intermediate valence properties of MnSi.

\section*{Acknowledgment}
We thank D. Aoki for the experimental helps and B. F\r{a}k and C. Meingast for fruitful discussions and comments.
This work has been partly financially supported by the French ANR programs ICENET, NEMSiCOM and ECCE.

\end{document}